\documentclass[fleqn,usenatbib]{mnras}

\usepackage{newtxtext,newtxmath}

\usepackage[T1]{fontenc}
\usepackage{ae,aecompl}

\usepackage{graphicx}	
\usepackage{amsmath}	
\usepackage{amssymb}	
\usepackage{enumerate}
\usepackage{gensymb}
\usepackage{textcomp}
\usepackage{float}

\title[The jet/wind outflow in Centaurus~A]{The jet/wind outflow in Centaurus~A: a local laboratory for AGN feedback}

\author[B. McKinley et al.]
{B.~McKinley,$^{1,2,3}$\thanks{E-mail:ben.mckinley@curtin.edu.au}
S.~J.~Tingay,$^{1,4}$
E.~Carretti,$^{5}$
S.~Ellis,$^{6}$
J.~Bland-Hawthorn,$^{7}$
\newauthor
R.~Morganti,$^{8,9}$
J.~Line,$^{3,2}$
M.~McDonald,$^{10}$
S.~Veilleux,$^{11}$
R.~Wahl~Olsen,$^{12}$
\newauthor
M.~Sidonio,$^{13}$
R.~Ekers,$^{14,1}$
A.~R.~Offringa,$^{8}$
P.~Procopio,$^{3,2}$
B.~Pindor,$^{3,2}$
\newauthor
R.~B.~Wayth,$^{1,2}$
N.~Hurley-Walker,$^{1}$
G.~Bernardi,$^{15,16}$
B.M.~Gaensler,$^{17,2}$
\newauthor
M.~Haverkorn,$^{18}$
M.~Kesteven,$^{14}$
S.~Poppi,$^{5}$
L.~Staveley-Smith$^{19,2}$
\\
$^{1}$International Centre for Radio Astronomy Research, Curtin University, Bentley, WA 6102, Australia\\
$^{2}$ARC Centre of Excellence for All-sky Astrophysics (CAASTRO), Curtin University, Bentley, WA 6102, Australia\\
$^{3}$School of Physics, The University of Melbourne, Parkville, VIC 3010, Australia\\
$^{4}$Istituto di Radioastronomia, Istituto Nazionale di Astrofisica, Via Gobetti 40129, Bologna, Italy\\
$^{5}$INAF Osservatorio Astronomico di Cagliari, Via della Scienza 5, 09047 Selargius (CA), Italy\\
$^{6}$Australian Astronomical Observatory, PO Box 915, North Ryde, NSW 1670, Australia \\
$^{7}$Sydney Institute for Astronomy, University of Sydney, NSW 2006, Australia \\
$^{8}$Netherlands Institute for Radio Astronomy (ASTRON), Postbus 2, 7990 AA Dwingeloo, The Netherlands\\
$^{9}$Kapteyn Astronomical Institute, University of Groningen, Postbus 800, 9700 AV Groningen, The Netherlands\\
$^{10}$Kavli Institute for Astrophysics and Space Research, MIT, Cambridge, MA 02139, USA \\
$^{11}$Department of Astronomy, Joint Space-Science Institute, University of Maryland, College Park, MD, 20742, USA \\
$^{12}$High View Observatory, Auckland 0612, New Zealand \\
$^{13}$Terroux Observatory, Canberra, ACT 2612, Australia \\
$^{14}$CSIRO Astronomy and Space Science, PO Box 76, Epping, NSW 1710, Australia\\
$^{15}$Square Kilometre Array South Africa (SKA SA), Park Road, Pinelands 7405, South Africa \\
$^{16}$Department of Physics and Electronics, Rhodes University, P.O. Box 94, Grahamstown, 6140, South Africa \\
$^{17}$Dunlap Institute for Astronomy and Astrophysics, University of Toronto, 50 St. George St, Toronto, ON M5S 3H4, Canada \\
$^{18}$Department of Astrophysics/IMAPP, Radboud University Nijmegen, PO Box 9010, 6500 GL Nijmegen, the Netherlands \\ 
$^{19}$International Centre for Radio Astronomy Research, University of Western Australia, Crawley, WA 6009, Australia 
}

\date{Accepted XXX. Received YYY; in original form ZZZ}

\pubyear{2017}

\begin{document}
\label{firstpage}
\pagerange{\pageref{firstpage}--\pageref{lastpage}}
\maketitle


\begin{abstract}
We present new radio and optical images of the nearest radio galaxy Centaurus~A and its host galaxy NGC~5128. We focus our investigation on the northern transition region, where energy is transported from the $\sim$5~kpc ($\sim$5~arcmin) scales of the Northern Inner Lobe (NIL) to the $\sim$30~kpc ($\sim$30~arcmin) scales of the Northern Middle Lobe (NML). Our Murchison Widefield Array observations at 154~MHz and our Parkes radio telescope observations at 2.3~GHz show diffuse radio emission connecting the NIL to the NML, in agreement with previous Australia Telescope Compact Array observations at 1.4~GHz. Comparison of these radio data with our widefield optical emission line images show the relationship between the NML radio emission and the ionised filaments that extend north from the NIL, and reveal a new ionised filament to the east, possibly associated with a galactic wind. Our deep optical images show clear evidence for a bipolar outflow from the central galaxy extending to intermediate scales, despite the non-detection of a southern radio counterpart to the NML. Thus, our observational overview of Centaurus~A reveals a number of features proposed to be associated with AGN feedback mechanisms, often cited as likely to have significant effects in galaxy evolution models. As one of the closest galaxies to us, Centaurus~A therefore provides a unique laboratory to examine feedback mechanisms in detail.\\
\end{abstract}



\clearpage
\section{Introduction}

Centaurus A (Cen~A), hosted in the galaxy NGC~5128, is recognised as the nearest classical radio galaxy, of Fanaroff-Riley type I (FR-I: \citealt{FR1974}) at a distance of 3.8$\pm$0.1 Mpc \citep{har10}.  Cen~A was the first radio source to be identified as extragalactic \citep{bol49} and (along with NGC 5128) has been studied intensively since, across the entire accessible electromagnetic spectrum and over eight orders of magnitude in spatial scale (see the proceedings of ``The Many Faces of Centaurus A'' for an overview: \citet{MFCA}).

As the closest radio galaxy, Cen~A affords an unprecedented opportunity to study the transport of energy from the central engine in the active galactic nucleus (AGN), powered by a 55-million solar mass black hole in this case \citep{cappellari2009,neu10}, to intergalactic medium scales well outside the host galaxy. Such systems are clearly excellent laboratories for studying the effects of AGN feedback \citep{cro06,fabian2012}. In the case of Cen~A, there is evidence of jet interactions with the environment within the galaxy \citep{tin09,wyk15} and of complex interactions with gas, both atomic \citep{mou00,oos2005,santoro2015a,santoro2015b,santoro2016}, and molecular, well beyond the galaxy itself. 

On scales of the AGN (parsec and below), Cen~A has been studied extensively over many years \citep{mei89,tin94,jon96,kel97,tin98,fuj00,tin01,tin01b,hor06,mul11,mul14}, revealing a collimated sub-pc-scale jet that evolves on timescales of months to years. This jet feeds highly complex structures on scales of kpc that have been documented using instruments like the Australia Telescope Compact Array \citep{fea11,morganti1999}, Very Large Array \citep{bur83,cla92,nef15} and GMRT \citep{wyk14}. On even larger scales, extended radio emission extends over many hundreds of kpc, imaged with interferometers \citep{mck13}, the Parkes radio telescope \citep{jun93} and other single dishes \citep{com97}.

On the scales of the inner lobes, the radio structure is symmetric around the AGN. The sub-pc-scale jet is also symmetrical (taking into account relativistic beaming), with double-sided jets aligned with the jets feeding the inner lobes. However, on larger spatial scales, Cen~A shows striking asymmetries. The so-called Northern Middle Lobe (NML) appears to be some form of extension to the Northern Inner Lobe (NIL), but the connection is unclear. Moreover, there is no obvious southern counterpart to the NML on scales larger than the inner lobes.

The study of Cen~A on a range of spatial scales, and across the electromagnetic spectrum, is useful in terms of tracing the transport of energy from the central black hole system to larger scales. As such, in this paper we present new radio data, primarily focusing on the intermediate and large spatial scale structure in Cen~A, at low frequencies from the Murchison Widefield Array (MWA) and at 2.3~GHz from the Parkes S-PASS survey. The radio spectral index can provide some insights into how energy is being transported and dissipated within a radio galaxy and so we use spectral index maps to investigate the physical processes at work within the lobes of Cen~A. We use the convention $S\propto\nu^{\alpha}$ to define the spectral index $\alpha$, where $S$ is the source flux density and $\nu$ is frequency.

One of the major challenges in studying Cen~A at radio frequencies is the fact that the radio source has complex structure on all angular scales, including features spanning several degrees across the sky.  This presents a major challenge for observations with traditional interferometers possessing limited short baselines and sub square degree fields of view.  Imaging artefacts have consistently held back many previous studies. Two advantages of the MWA in this respect are: 1) the immense number of very short baselines, providing fantastic sensitivity to large scale structures; and 2) the extraordinarily large field of view (hundreds of square degrees) that accommodates Cen~A within a single observation. We exploit these advantages to produce the best low frequency images (in some respects the best at any frequency) of Cen~A yet. Similarly, single-dish observations with Parkes, used for the S-PASS survey, are sensitive to the largest angular scales on the sky and despite having poorer angular resolution than the MWA, the Parkes images have less imaging artefacts, which increases their dynamic range.

Additional clues as to the energy transport mechanisms at play in the northern transition region can also be gleaned from optical data, in particular from the filamentary structures that result from photoionisation of gas by starburst or AGN activity. The bright optical filaments north of NGC~5128 were originally discovered by \citet{blanco1975} and \citet{peterson1975}. There have been many competing models discussed in the literature to explain their formation and structure (see \citealt{duf1978}, \citealt{graham1981}, \citealt{graham1983}, \citealt{sutherland1993}, \citealt{morganti1991}, \citealt{morganti1999}, \citealt{santoro2015a}, \citealt{rej2002}, \citealt{nef15B}). In this paper we take advantage of the narrowband sensitivity and wide field of view of the Maryland-Magellan Tunable Filter (MMTF; \citealt{MMTF}) on the Inamori Magellan Areal Camera and Spectrograph (IMACS) at the Magellan Telescope to examine the relationship between the optical emission line filaments and the radio emission of Cen~A.

As the fifth brightest galaxy in the sky \citep{israel1998}, NGC~5128 is also a popular target for amateur astronomers. The relatively large angular extent of the galaxy, including emission from stars in an extended halo of shell-like structures \citep{malin1983,haynes1983} and optical extensions spanning tens of arcmins \citep{cannon1981}, make it well-suited to smaller telescopes with large fields of view. In this vein, we include a deep optical image constructed from data acquired using two amateur set-ups. which captures the shells and optical extensions in unprecedented detail, allowing us to carefully compare the optical continuum morphology to the radio and emission-line structures.

In subsequent sections, we describe the structure of Cen~A and introduce the nomenclature used throughout this paper (Section~\ref{sec:nomenclature}). We then describe our new observations at radio and optical wavelengths that are the main focus of this paper, including relevant details of the data reduction processes involved (Section 3). In Section 4 we present our analysis and describe the large-scale spectral properties of the radio component in detail. We then compare our new data with previous radio datasets and results, to provide new insights into some of the outstanding questions regarding Cen~A, for example the structure of the NML and the absence of a southern counterpart, the nature of energy transport from the inner lobes to the NML and mechanisms responsible for the formation of the prominent optical filaments at the base of the NML (Section 5). We provide our conclusions and recommendations for further study in Section 6. 

\section{Cen~A structure and nomenclature}
\label{sec:nomenclature}

\begin{figure*}
\centering 
\includegraphics[clip,trim=110 70 267 63,width=1.0\textwidth,angle=0]{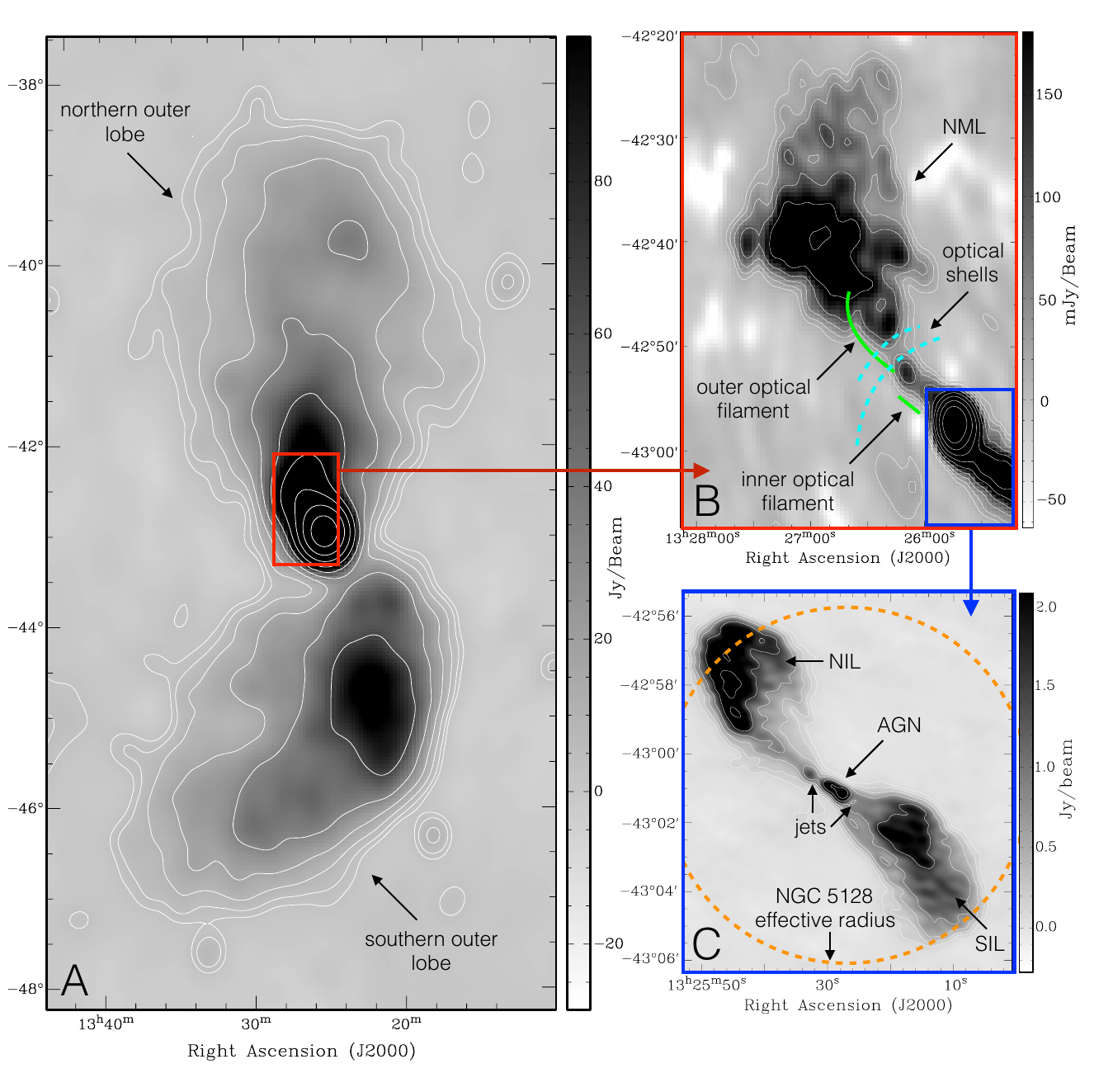}
\caption{Overview of Cen~A structure and nomenclature. Panel A: Large scales. This image showing the outer lobes of Cen~A has been redrawn from the published data of \citet{mck13} with contours from 2 to 1024 Jy/beam, incrementing in a geometric progression with a common factor of 2. The image was taken with the MWA 32-tile prototype at a frequency of 118~MHz and an angular resolution of 25~arcmin. Panel B: Intermediate scales. This image showing the NML has been redrawn from the published data of \citet{morganti1999} with contours from 12.5 to 200~mJy/beam, incrementing in a geometric progression with a common factor of 2. The image was taken with ATCA at a frequency of 1.4~GHz and an angular resolution of approximately 2~arcmin. The image has also been annotated with the positions of the outer and inner optical filaments in solid green (estimated from the data presented in this paper) and the positions of the two outermost optical shells (estimated from figure~2 of \citet{malin1983}, where they are labeled 17 and 18) in dashed cyan. Panel C: Small scales. This image showing the inner lobes has been redrawn from the published data of \citet{condon1996} with contours from 0.1 to 3.2 Jy/beam incrementing in a geometric progression with a common factor of 2. The image was taken with the VLA at a frequency of 1.4~GHz and an angular resolution of 18~arcsec. The dashed orange circle indicates the effective radius of NGC~5128 of 305~arcsec \citep{dufour1979}. A colour version of this figure is available in the online article.}
\label{CenA_overview}
\end{figure*}

Due to the extensive study of Cen~A, there exist commonly-used names for many of its features, as observed across a wide range of wavelengths. In Fig.\ref{CenA_overview}, we show an overview of the Cen~A structure at large, intermediate and small scales, as observed in the radio. In this paper we categorise the features of Cen~A as follows: \\ \\
1. Large scales (degrees/$\sim$10$^{2}$~kpc), characterised by the radio emission of the outer lobes. For observational examples see \citet{jun93}, \citet{fea11}, \citet{mck13}, \citet{israel1998}, \citet{alvarez2000}, \citet{osullivan2013}. The outer lobes are shown in Fig.~\ref{CenA_overview}, panel A. \\
2. Intermediate scales (arcmins/kpc), characterised in the radio by the NML \citep{morganti1999}, the southern transition region \citep{osullivan2013} and the surrounding gas clouds \citep{HI1994,oos2005,morganti2016,salome2016} and in the optical by the inner and outer optical filaments \citep{blanco1975} and the halo stars and shells \citep{malin1983}. The radio emission from the NML is shown in Fig.~\ref{CenA_overview}, panel B.\\
3. Small scales (arcsecs, pc/sub-pc), characterised in the radio by the inner lobes (NIL and SIL; \citealt{cla92}) and the radio jets \citep{tin98} and in the optical by the host galaxy NGC~5128 \citep{graham1979,dufour1979}. The radio emission of the NIL and SIL are shown in Fig.~\ref{CenA_overview}, panel C.\\

Our new observations are well-suited to studying the intermediate scales as defined above and so it is at these scales that we focus our attention in this paper.

\section{Observations and data reduction}

\subsection{MWA Observations and data reduction}

The MWA \citep{tingay,bowman2013} is a low-frequency radio interferometer array located at the Murchison Radio-astronomy Observatory in Western Australia. The telescope is made up of 128 antenna tiles, each containing 16 crossed-dipole antennas above a conducting ground plane. The tiles are pointed electronically using analogue beamformers. 

The observations of Cen~A used in this paper were undertaken as part of the Galactic and Extragalactic All-sky MWA (GLEAM) survey \citep{wayth2015,HW2017} on 2014 June 10. Due to the high signal-to-noise ratio (SNR) afforded by the brightness of Cen~A and the exceptional instantaneous $(u,v)$ coverage of the MWA, we were able to produce our image from a single 112~s snapshot observation beginning at UTC 12:42:24. In line with the observing strategy of this phase of the GLEAM survey, our snapshot was part of a series of drift-scan observations at a fixed declination of $-40$\degr. The centre frequency of this particular observation was approximately 154~MHz and covered the MWA instantaneous bandwidth of 30.72~MHz. We used a correlator mode that produced data at a frequency resolution of 40~kHz and a time resolution of 0.5~s.

Initial calibration was performed using the MWA Real Time System (RTS; \citealt{mitch2008}, Mitchell et al. 2017, in prep.) operating in an offline mode on Galaxy, a supercomputer located at the Pawsey Supercomputing Centre in Western Australia. We first constructed a sky model from archival multi-wavelength data, cross-matched using the Positional Update and Matching Algorithm (PUMA; \citealt{line2017}). The catalogues included in the cross-match were: 74~MHz Very Large Array Low Frequency Sky Survey redux (VLSSr; \citealt{lane2012}), the 843~MHz Sydney University Molonglo Sky Survey (SUMSS; \citealt{mauch2003}), the 1.4~GHz NRAO VLA Sky Survey (NVSS; \citealt{condon1998}) and GLEAM (Internal Data Release 2; \citealt{HW2017}). The sky-model generated from the cross-match contained the brightest 1000 sources for our pointing on the sky, taking into account the MWA primary beam. To this point-source sky model, we added a Gaussian model of the Cen~A inner lobes, constructed by performing source extraction with PyBDSM \citep{pybdsm} on the VLA 1.4~GHz image of the inner lobes \citep{condon1996}, with the flux density scaled assuming a spectral index of $-0.5$ \citep{cla92}.

Using our sky model as an input to the RTS, the calibration was performed on a per-channel basis at a cadence of 8~s. The output from this process was a calibrated uvfits dataset, which was then imaged with \textsc{wsclean} \citep{wsclean}, using a relatively high clean threshold, Briggs weighting with robust parameter zero and using the multiscale imaging option. Several rounds of self-calibration were performed using \textsc{calibrate} (the calibration tool developed and applied to MWA data by \citealt{offringa2016}), using the multiscale model produced by \textsc{wsclean}. The image was primary beam corrected with the primary beam model of \citet{sutinjo2014}, using \textsc{pbcorrect} \citep{offringa2016}. Source finding was performed on the final image and the resulting source catalogue was cross-matched with the catalogues used to generate the original calibration model using PUMA. The cross match revealed a positional offset between the original model and the catalogue of sources produced from the primary-beam-corrected image. This was corrected by adjusting the FITS header of the image. The PUMA outputs from this comparison also enabled us to check the flux-density scale of our image, which showed that our source flux densities match well with the expected flux densities at 150~MHz as predicted by PUMA.

\subsection{Parkes Observations and data reduction}

The S-band Polarization All Sky Survey (S-PASS) has mapped the entire southern sky at Dec $<-1.0$\degr \ in total intensity and linear polarisation with the 64-m Parkes Radio Telescope at a frequency of 2300~MHz. A full description of S-PASS is given in \citet{carretti2013} and \citet{carretti2010}, here a summary of the main details is reported. The Parkes S-band receiver was used; a receiver package with system temperature T$_{sys}\approx$20~K, beamwidth FWHM=8.9~arcmin, and a circular polarisation front-end, that is an ideal setup for linear polarisation observations with single-dish telescopes. Data were collected with the digital correlator Digital Filter Banks mark 3 (DFB3) recording full Stokes information (autocorrelation products of the two circular polarisations RR* and LL*, and the real and imaginary parts of their complex cross-product RL*). Here we focus on the description of the Stokes~I observations, since only these were used in this work.

Correlator data were initially binned in 8~MHz channels for calibration and radio frequency interference (RFI) flagging purposes. The sources PKS~B1934-638 and PKS~B0407-658 were used for flux density calibration as primary and secondary calibrators, respectively. The flux scale model \citet{reynolds1994} was assumed for PKS~B1934-638. After RFI flagging, the useful band covered the ranges 2176-2216~MHz and 2256-2400~MHz. All useful 8~MHz bins were eventually binned together into one channel for an effective central frequency of 2307~MHz and 184~MHz bandwidth. Observing was carried out as long azimuth scans taken at the elevation of the south celestial pole at Parkes covering the entire Dec range at each scan ($-90$\degr \ to $-1$\degr). The scans were taken in the east and the west to attain an effective basket weaving pattern and realise absolute polarisation calibration of the data. Full details of the scanning strategy and map-making technique are given in \citet{carretti2010} and in the forthcoming S-PASS survey paper (Carretti et al., in preparation). Final maps were convolved to a beam of FWHM=10.75~arcmin. For this analysis a $30^\circ \times 20^\circ$ map in zenithal equidistant (ARC) projection centred at Galactic coordinates ($l$, $b$) = (309.50$^\circ$, 19.50$^\circ$) was extracted. That included Cen~A and an outer region to estimate the background/foreground emission at the Cen~A position. Stokes I, Q and U sensitivity is better than 1.0~mJy beam$^{-1}$ in the area covered by S-PASS. The Stokes~I confusion limit has been measured at 13~mJy beam$^{-1}$ directly from the S-PASS map \citep{meyers2017} and estimated to be much lower in polarisation (average polarisation fraction of S-PASS compact sources is $\sim$2\% according to \citealt{lamee2016}). 

\subsection{Optical emission line observations and data reduction}
\label{sec:MMTF_section}

Narrow band imaging observations were made on 2009 February 25 using the MMTF \citep{MMTF} on IMACS at the Magellan Telescope. The MMTF has a very narrow bandpass ($\sim$5 to 12~\AA), which can be tuned to any wavelength over the range 5000 to 9200~\AA\ \citep{MMTF}. This instrument is ideal for detecting emission-line filaments in external galaxies as it combines the excellent image quality at Magellan with the wide field of view of IMACS.  On 2009 February 25, we observed Cen~A at H$\alpha$ ($\lambda$ = 6563~\AA) and [N\textsc{II}] ($\lambda$ = 6584~\AA) for a total of 20~min each and in the R band for 30~s. The typical image quality for these exposures was 0.7 $\pm$ 0.2~arcsec. Note that narrow band tunable filter images have a phase effect such that the central wavelength of the passband changes as a function of radius from the centre of the field (see \citealt{JSB}). The central wavelengths for both the H$\alpha$ and [NII] images were set at the location of the outer optical filament, since the H$\alpha$ structure of the inner filament is already well-established \citep{hamer2015}.

The data were reduced using the MMTF data reduction pipeline\footnote{http://www.astro.umd.edu/$\sim$veilleux/mmtf/datared.html}, following the method described by \citet{mcdonald2010}. Further details of the data reduction methods can be found in \citet{MMTF} and \citet{JSB}. As per \citet{mcdonald2010}, the spectrophotometric standards used for calibration come from \citep{oke1990} and \citep{hamuy1992,hamuy1994} and the resulting absolute photometric calibration error is $\sim$15\%, as is typical for tunable filters and spectrographs such as the MMTF.


\subsection{Deep optical observations and data reduction}
\label{sec:optical_obs}

Our deep optical imaging was undertaken in two separate observing campaigns in Auckland, New Zealand and in Wiruna, Australia. NGC~5128 was observed from Auckland over 43 nights between 2013~February and 2013~May, using a homebuilt 10~inch f/4 Serrurier Truss Newtonian on a Losmandy G-11 mount, QSI683wsg-8 CCD camera with Astrodon LRGB E-Series Gen~2 filters and a Lodestar guider. Imaging was performed using the LRGB technique with total exposure times for the luminance, red, green and blue channels of 90, 10, 10 and 10~h, respectively. Image integration and processing were performed using \textsc{PixInsight~1.8}. Image calibration was performed against an extensive set of calibration frames consisting of approximately 400 dark frames, 500 bias frames and 300 flat-field frames in each channel (L, R, G and B). All calibration frames were obtained at or near the typical operating temperature of the CCD camera ($-25$\degr C). All light frames were calibrated using the automatic dark/bias scaling feature of \textsc{PixInsight~1.8} to compensate for any temperature differences.

NGC~5128 was observed from Wiruna for 19.5~h over three nights using a 152~mm F7.5 apochromatic refractor with 4" field flattener, FLI ProLine11002 CCD camera and Astronomik LRGB filters. The data were collected from very dark skies and under good seeing conditions, 1000~m above sea level. Total exposure times were 15, 1.5, 1.5 and 1.5~h for the L, R, G, B channels, respectively. The data were acquired at $-35$\degr C chip temperature and calibrated with multiple dark and flat frames. Frames were all median-combined and pre- and post-processing was performed using the \textsc{astroart} image processing package.

\section{Results and Analysis}

\subsection{MWA image}

\begin{figure*}
\centering 
\includegraphics[clip,trim=140 110 130 143,width=1.0\textwidth,angle=0]{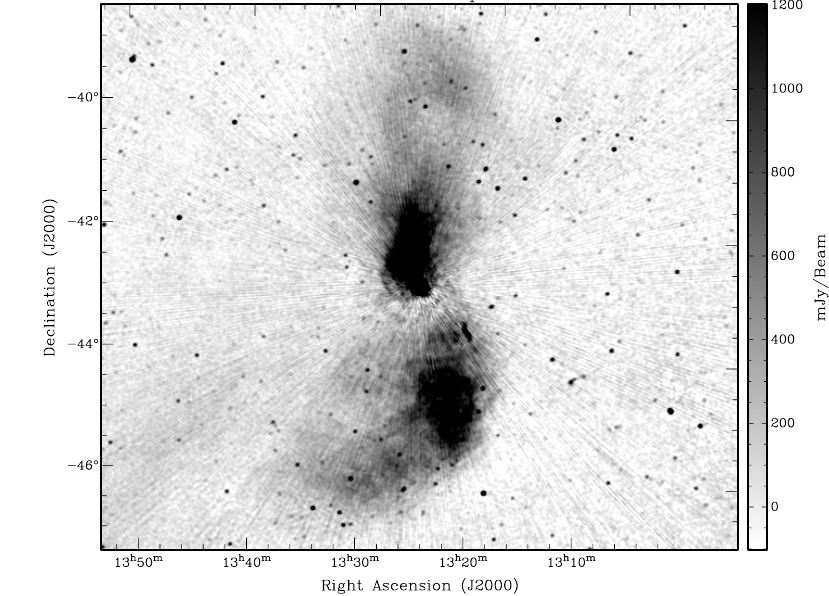}
\caption{Cen~A at 154~MHz with the Murchison Widefield Array. The image is shown on a linear scale between $-0.1$ and 1.2~Jy/beam and has an angular resolution of 3~arcmin.}
\label{CenA_zoom_nocont}
\end{figure*}

The MWA image at 154~MHz is shown in Fig.~\ref{CenA_zoom_nocont}. This is a cropped version of the full image, which covers a much larger field of view, spanning approximately 30\degr\ out to the first null of the primary beam. The image has been restored with a Gaussian beam of width 3.00$\times$2.76~arcmin with major axis position angle of 112.07\degr. The peak brightness in the image is 358~Jy/beam at position RA~(J2000)~13\textsuperscript{h}25\textsuperscript{m}42.5\textsuperscript{s}, Dec~(J2000)~$-42$\degr 57$'$31.6$''$, corresponding to a point in the NIL. The rms noise level of the image, away from any obvious radio sources, is approximately 50~mJy/beam, giving a dynamic range of approximately 7000. The noise is dominated by faint radial artefacts originating from the bright core of Cen~A. These artefacts are due to small calibration errors that are amplified by the extreme brightness of the inner lobes of Cen~A.

\subsection{Parkes images}

\begin{figure*}
\centering 
\includegraphics[clip,trim=145 100 125 147,width=1.0\textwidth,angle=0]{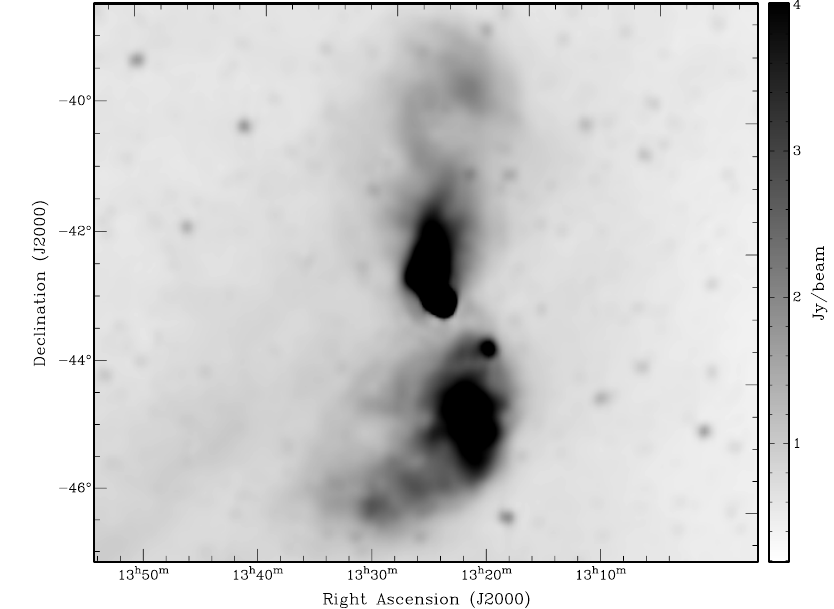}
\caption{Cen~A at 2307~MHz with the Parkes radio telescope. The image is shown on a linear scale between 0.2 and 4~Jy/beam and has an angular resolution of 11~arcmin.}
\label{parkes_zoom}
\end{figure*}

The Parkes S-PASS image at 2.3~GHz is shown in Fig.~\ref{parkes_zoom}. This is a cropped version of the full image, which covers a much larger field of view, spanning the entire southern sky. The size of the single-dish beam is 10.8~arcmin. The peak brightness in the image is 144~Jy/beam at position RA~(J2000)~13\textsuperscript{h}25\textsuperscript{m}36\textsuperscript{s}, Dec~(J2000)~$-42$\degr 59$'$37$''$. The rms noise level of the image, away from any obvious radio sources, is approximately 15~mJy/beam, giving a dynamic range of approximately 9600.

\subsection{Spectral index between 154~MHz and 2.3~GHz}
\label{sec:spectral_index}

\begin{figure*}
     \begin{minipage}[r]{1.2\columnwidth}
         \raggedright
         \includegraphics[clip,trim=260 70 190 95,width=1.33\linewidth]{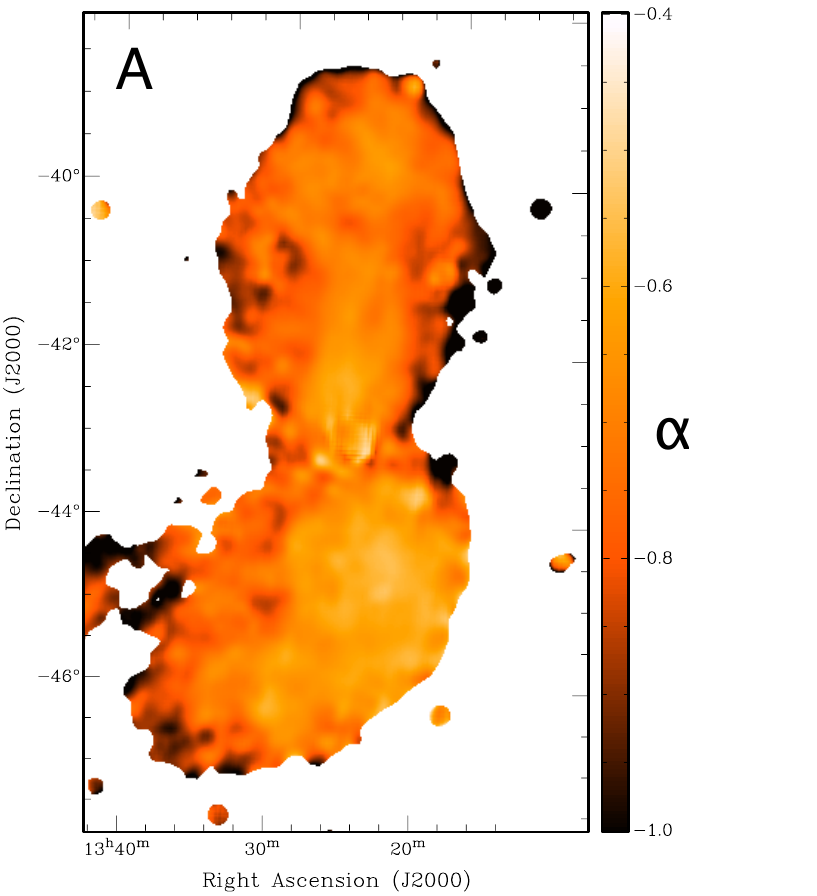}
     \end{minipage} 
     \hfill\begin{minipage}[l]{0.76\columnwidth}
         \hfill\includegraphics[clip,trim=280 140 300 183,width=1.1\linewidth]{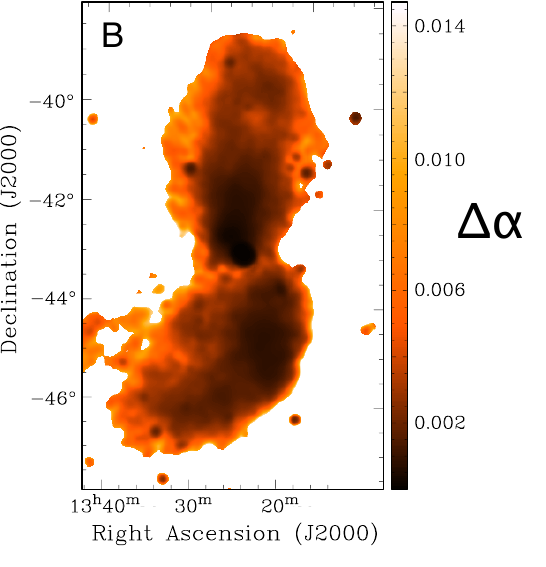} 
         \raggedleft
         \newline
         \includegraphics[clip,trim=280 140 300 183,width=1.1\linewidth]{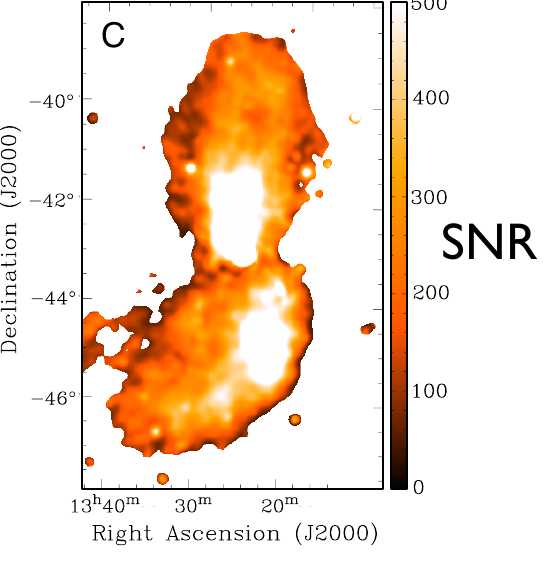}
     \end{minipage}

\caption{Spectral index analysis of Cen A between 154~MHz and 2.3~GHz. Panel A shows the spectral index $\alpha$ as defined by $S\propto\nu^{\alpha}$. Panel B shows the error in the spectral index, computed by propagation of errors, using the background rms noise of the images as the error estimate across the entire image. Panel C shows the signal to noise ratio computed by taking the ratio of the images in panels A and B. A colour version of this figure is available in the online article.}
\label{spec_index_figure}
\end{figure*}

We constructed a spectral index map between 154~MHz and 2307~MHz, which is shown in Fig.~\ref{spec_index_figure}, panel A. The spectral index was calculated using:  

\begin{equation} 
\alpha=\frac{\rm{\log}(S_{154}/S_{2307})}{\rm{\log}(154/2307)},
\label{alpha2}
\end{equation}
where $S_{154}$ and $S_{2307}$ are the flux density values of each pixel in the 154~MHz and 2307~MHz images, respectively.

For an accurate representation of the spectral index it is important to ensure that both images contain information on the same angular scales. The smallest baseline of the MWA is approximately 7.7~m, allowing the interferometer to sample angular scales of up to 14.5\degr \ at zenith. This is more than sufficient to fully sample the emission from Cen~A, which has a maximum extent of approximately 8\degr. The single-dish S-PASS image of course contains larger spatial scales, including the average or `zero-baseline' value. 

The following procedure was used to effectively match the spatial information contained in both images. We first cropped the wide-field MWA image to an 865~$\times$~865 pixel sub-image covering an area of 10\degr$\times$10\degr\ centred on Cen~A. This image was used as a template to regrid the S-PASS image to obtain a 2307~MHz image covering the same area as the cropped MWA image and having the same image and pixel sizes. We then took the FFT of the resultant S-PASS image, set the central `zero spacing' pixel in the Fourier plane to zero, and took the inverse FFT to obtain an image with the mean value removed, thus obtaining a 10\degr$\times$10\degr\ image containing the same information on large spatial scales as the MWA image. The higher-angular-resolution MWA image was smoothed to the S-PASS angular resolution of 10.75~arcmin. The spectral index image was then calculated using Equation~(\ref{alpha2}).

Panels B and C of Fig. \ref{spec_index_figure} show the error in the spectral index and the SNR in the spectral index map, respectively. The error in the spectral index was calculated by propagation of the error in the individual images, which is taken to be the rms noise in a region outside of Cen~A. We can see that the SNR is high throughout the lobes (the lowest value is around 20 at the edges of the lobes where the emission is faintest), giving us confidence that our spectral variations on the order of 0.01 for $\alpha$ are real.

\begin{figure*}
\centering 
\includegraphics[clip,trim=70 120 10 100,width=0.82\textwidth,angle=270]{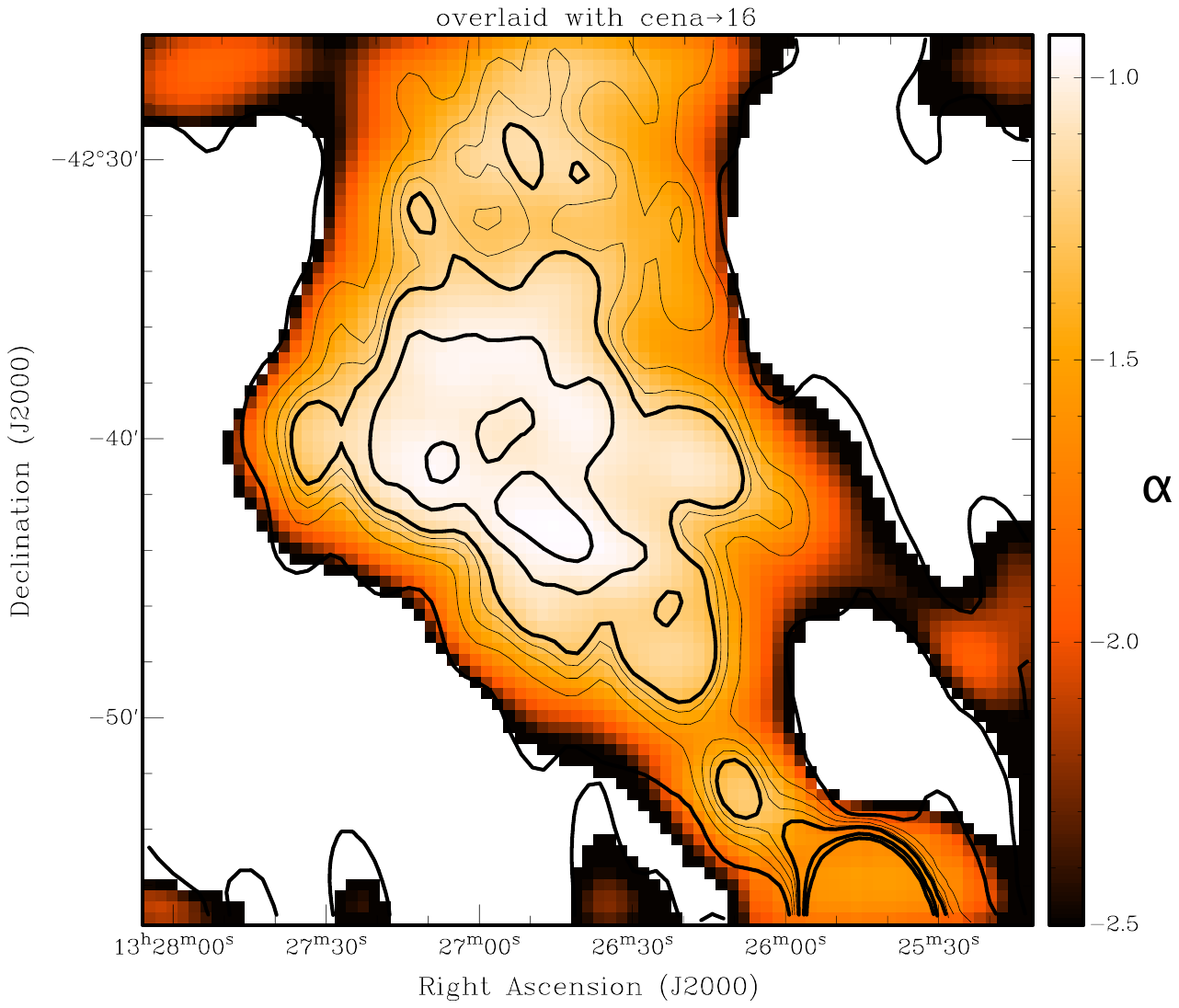}
\caption{Spectral index map of the Cen~A NML between 154~MHz and 1.4~GHz. Contours are from the 1.4~GHz image of \citet{morganti1999} at 0, 0.025, 0.05, 0.075, 0.1, 0.2, 0.3 Jy/beam. A colour version of this figure is available in the online article.}
\label{MWA_ATCA_SPEC}
\end{figure*}

To answer some of the outstanding questions related to the NML, we have also constructed a spectral index map between 1.4~GHz and 154~MHz, which is presented in Fig.~\ref{MWA_ATCA_SPEC} and discussed in Section~\ref{sec:mwa_atca_spec}.

\subsection{Optical emission line images of NML}

\begin{figure*}
\centering 
\includegraphics[clip,trim=30 130 30 312,width=1.0\textwidth,angle=0]{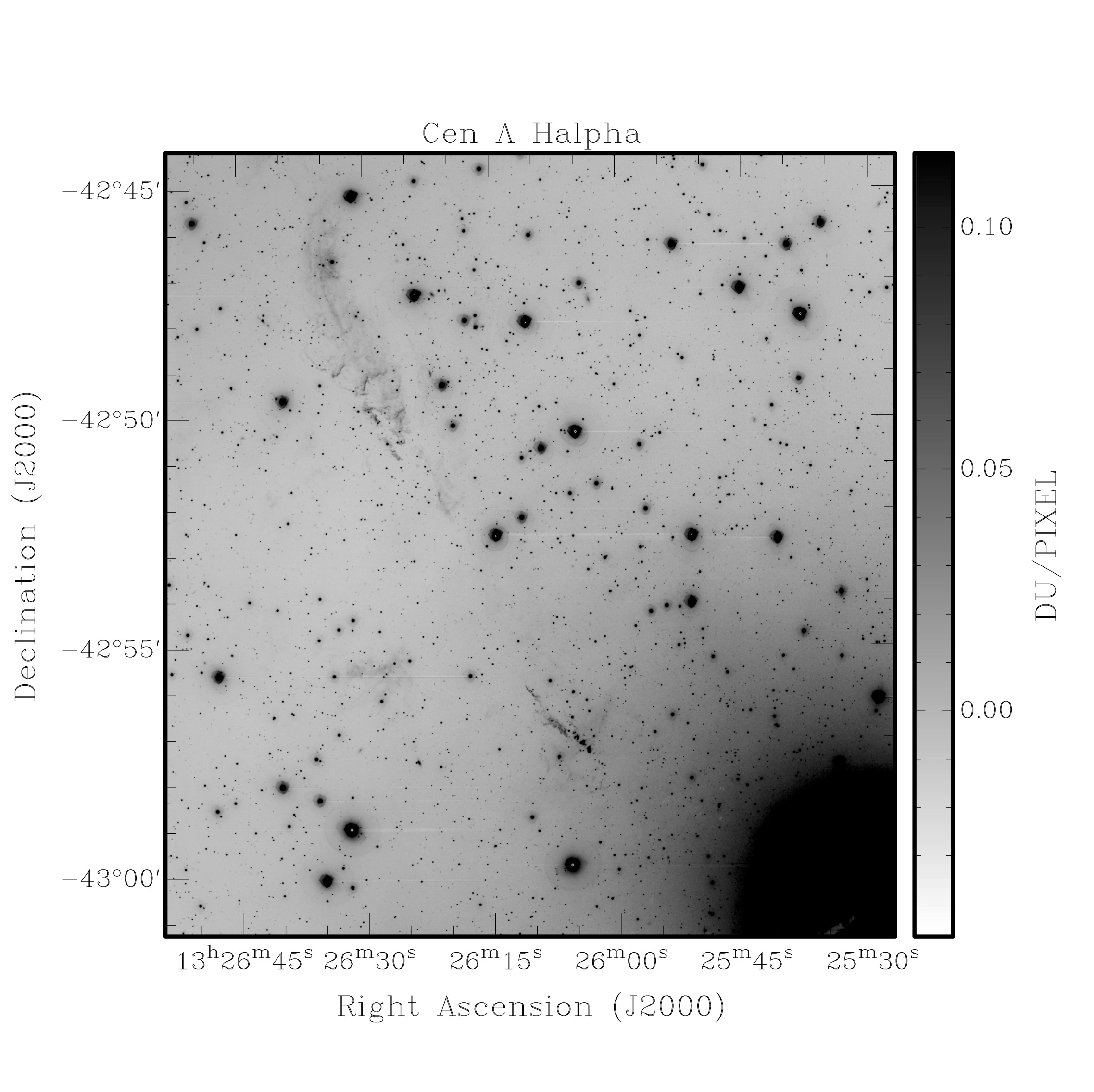}
\caption{H$\alpha$ ($\lambda$ = 6563~\AA) image obtained with MMTF showing the optical filaments in the northern transition region of Cen~A.}
\label{optical_emission_lines}
\end{figure*}

The H$\alpha$ ($\lambda$ = 6563~\AA) image obtained with MMTF is shown in Fig.~\ref{optical_emission_lines}. The image covers a 15$'$ patch of the northern transition region where the NIL meets the NML. The central wavelength of the MMTF was set at the location of the outer optical filament, which can be seen in the north-east quadrant of the image. As described by \citet{MMTF}, the MMTF is ideally suited to observations of faint emission-line objects embedded in bright optical continuum. This has allowed us to identify a previously unexamined filament, which can be seen in Fig.~\ref{optical_emission_lines} at position RA~(J2000)~13\textsuperscript{h}26\textsuperscript{m}30\textsuperscript{s}, Dec~(J2000)~$-42$\degr 56$'$ and is discussed further in subsequent sections. 

\subsection{Deep optical images of the NML}

\begin{figure*}
\centering 
\includegraphics[clip,trim=0 7 0 30,width=0.9\textwidth,angle=0]{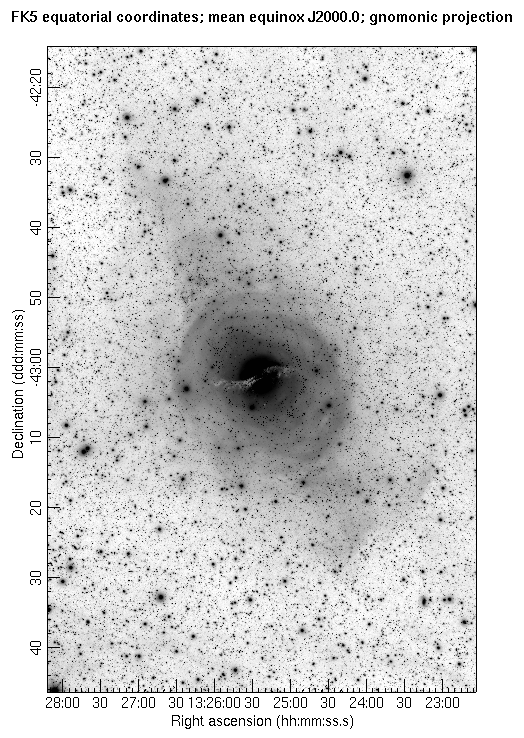}
\caption{Deep, widefield optical image of NGC~5128 using data obtained with a 254~mm Newtonian reflector and a 152~mm apochromatic refractor and using LRGB filters covering a wavelength range of 400-700~nm.}
\label{optical1}
\end{figure*}

The deep optical image produced using a combination of data from two amateur telescope setups, as described in Section~\ref{sec:optical_obs}, is shown in Fig.~\ref{optical1}. The image shows the prominent dust lane and halo stars forming shells \citep{malin1983,peng2002,haynes1983} and bipolar structures \citep{cannon1981,haynes1983} extending out into the northern and southern transition regions of Cen~A. The wide field of view of the telescopes and long exposure times afforded by a pair of dedicated and skilled amateur astronomers have produced an image where the shell structure is clearly visible without the need for any image post-processing, such as unsharp masking used in previous work, to accentuate the shells. The bipolar optical extensions are also clearly visible, allowing us to compare them in detail with our radio images, which we do in Section~\ref{sec:halo}.

\section{Discussion}

\subsection{The morphology and spectral properties of the outer lobes}

The large-scale morphology of the outer lobes of Cen~A revealed by the MWA at 3~arcmin resolution is consistent with previous studies at lower \citep{jun93} and higher \citep{fea11} angular resolutions and shows the prominent hook structure of the northern outer lobe and the more gradual curve of the southern outer lobe from west to east. As suggested by other authors, such as \citet{haynes1983} and based on the morphology of the radio lobes, we argue that the jets of the AGN may have once been oriented in a more N-S direction, excavating a pair of cocoons in the intergalactic medium (IGM) symmetrically north and south of the AGN and that since that time, the jets have precessed or been reoriented by some disruptive event. In that scenario, the current shape of both lobes would be the result of more recent outbursts, having an orientation the same as the current orientation of the inner jets with a position angle of 55\degr \ anti-clockwise from the north-south axis (\citealt{schreier1981}, \citealt{cla92}). This would result in the `S' shape of the radio galaxy that we see today, as new material hits the IGM on the east side of the cocoon in the north and the west side of the cocoon in the south. There is an asymmetry, however, in the radio morphology on intermediate and large scales. In the north there is the bright NML and a much straighter N-S oriented outer lobe (but still with a west to east protruding `hook'), while in the south there is no `southern middle lobe' and the outer lobe largely resembles the SIL in shape. We discuss this asymmetry further in Section~\ref{sec:HI}.

The spectral index map shown in Fig.~\ref{spec_index_figure} shows a remarkably uniform spectral index of between $-0.6$ and $-0.8$ across most of the outer lobes, with little variation with distance from the AGN along the jet axis or in the transverse direction across the lobes. As noted by many other authors, the spectral age of $\sim$30~Myr inferred for the lobes, based on synchrotron and inverse-Compton cooling times \citep{israel1998,alvarez2000,hardcastle2009,yang2012} must underestimate their true age due to dynamical arguments, and therefore particle re-acceleration is likely to be occurring (see e.g. \citealt{fea11}, \citealt{stawarz2013}, \citealt{stefan2013}, \citealt{wyk13}, \citealt{eilek2014}). Spatially-varying magnetic field strengths could also be a factor in the discrepancy between the dynamical and spectral ages of the outer lobes \citep{wyk15b}.  

The uniformity of the spectral index is evidence that either fresh electrons are being continually supplied from the AGN, or in-situ particle acceleration is taking place in the lobes (or most likely a combination of the two). In either case, energy is being supplied to the outer lobes from the AGN, otherwise even in-situ particle acceleration would cease in a relatively short amount of time \citep{wyk14,eilek2014}, causing the outer lobes to fade from view. In-situ particle acceleration effectively erases any information on the ages of the lobes that can be inferred from their spectra, leaving us to rely on dynamical arguments such as the sound crossing time (see \citealt{wyk14}) to provide constraints on the ages of the lobes. It is therefore possible that the outer lobes pre-date the merger event that occurred on order 120~Myr ago and that merger activity has had nothing to do with initiating AGN outbursts. Even though the outer lobes may in fact be very old, they are clearly not `dead' and must be being resupplied with energy, the most likely source of which is the AGN, via the inner lobes.

\subsection{Radio emission in the Northern Middle Lobe: existence of a large-scale `jet'}

One of the many unanswered questions about the radio lobes of Cen~A is the means of energy transport from the inner lobes to the outer lobes. Viable models of the source require that the giant outer lobes are much older (on the order 1~Gyr) than the radiative lifetimes of their constituent electrons. This is due primarily to the sound crossing time in the IGM and the time required to excavate the outer lobes \citep{eilek2014}. This indicates that ongoing, in-situ particle acceleration must be occurring within the lobes. In order to maintain the turbulence necessary for this in-situ particle reacceleration, models require that the outer lobes are resupplied with energy from the central galaxy, otherwise the turbulence would decay within 30~Myr \citep{eilek2014,nef15B,wyk14}. However, the means by which energy is transported from the inner lobes to the outer lobes are not well understood. To understand the physical mechanisms involved, we must understand the astrophysics of the so-called transition regions, between the inner and outer lobes of Cen~A, both in the north and the south.

The northern transition region, dominated by the NML, has been the subject of several studies at radio frequencies, the most relevant for this discussion being \citealt{morganti1999} (hereafter M99) and \citealt{nef15} (hereafter N15A). The NML is a radio-loud structure, the second brightest feature of Cen~A after the bright inner lobes, and there are several competing models to explain its morphology and its relationship to the inner and outer northern lobes. One set of models invokes a direct radio-loud connection between the NIL and the NML through a collimated jet (e.g. M99, \citealt{jun93}, \citealt{kraft2009}, \citealt{romero1996}, \citealt{gopal2010}). Other models do not require the presence of a large-scale jet, for example \citet{saxton2001} who model the NML as a buoyant bubble of plasma from a previous AGN outburst and \citet{crockett2012} who propose a weak cocoon-driven bow shock, rather than a shock generated directly by a jet. However, such models fail to explain some of the short-lived features of the NML that require a recent resupply of energy. N15A and \citealt{nef15B} (hereafter N15B) introduce a galactic wind, driven in part by starburst activity in NGC~5128, allowing the existence of the features of the NML without a distinctly collimated large-scale jet.

Australia Telescope Compact Array (ATCA) observations at 20~cm by M99 support the existence of a large-scale jet connecting the NIL and the NML, as their image clearly shows a linear structure between the two features. However, recent Very Large Array (VLA) observations at 90~cm by N15A have brought the existence of this feature once again into question. Their image shows a gap in the diffuse emission to the north-east of the NIL, almost exactly coincident with the M99 linear connection feature. N15A also claim to have detected a new diffuse emission feature to the north of the NIL (see their figure 3, upper panel).

\begin{figure*}
     \begin{minipage}[r]{1.0\columnwidth}
         \raggedright
         \includegraphics[clip,trim=200 70 210 110,width=1.05\linewidth]{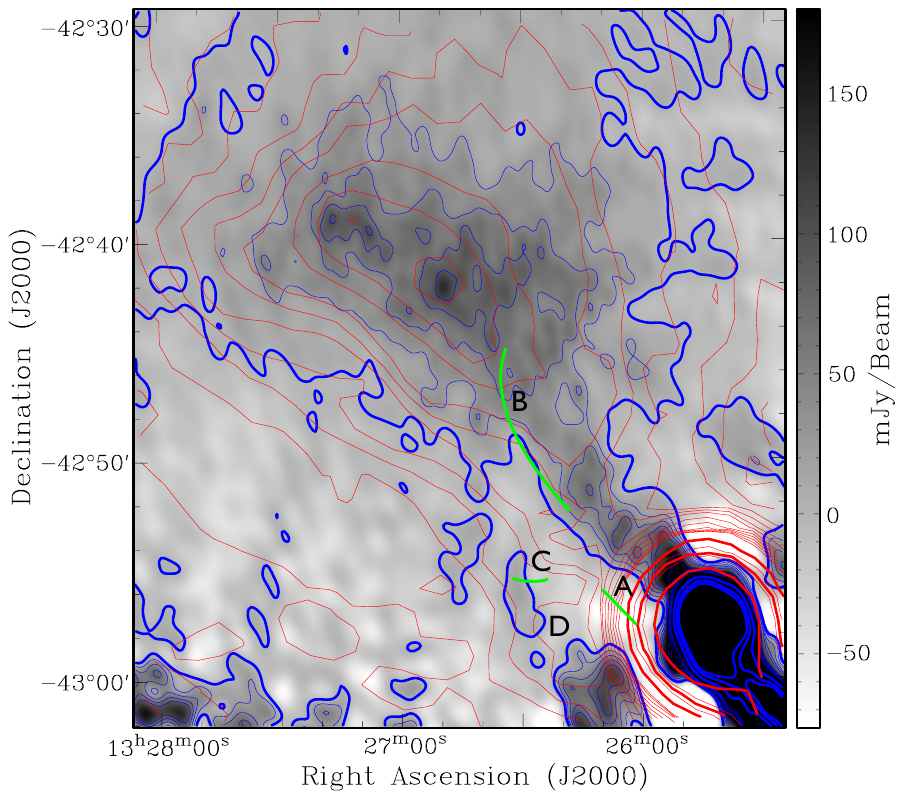}
     \end{minipage}
     \hfill\begin{minipage}[l]{1.0\columnwidth}
         \hfill\includegraphics[clip,trim=200 70 210 110,width=1.05                                                                                                                                                                                                                                  \linewidth]{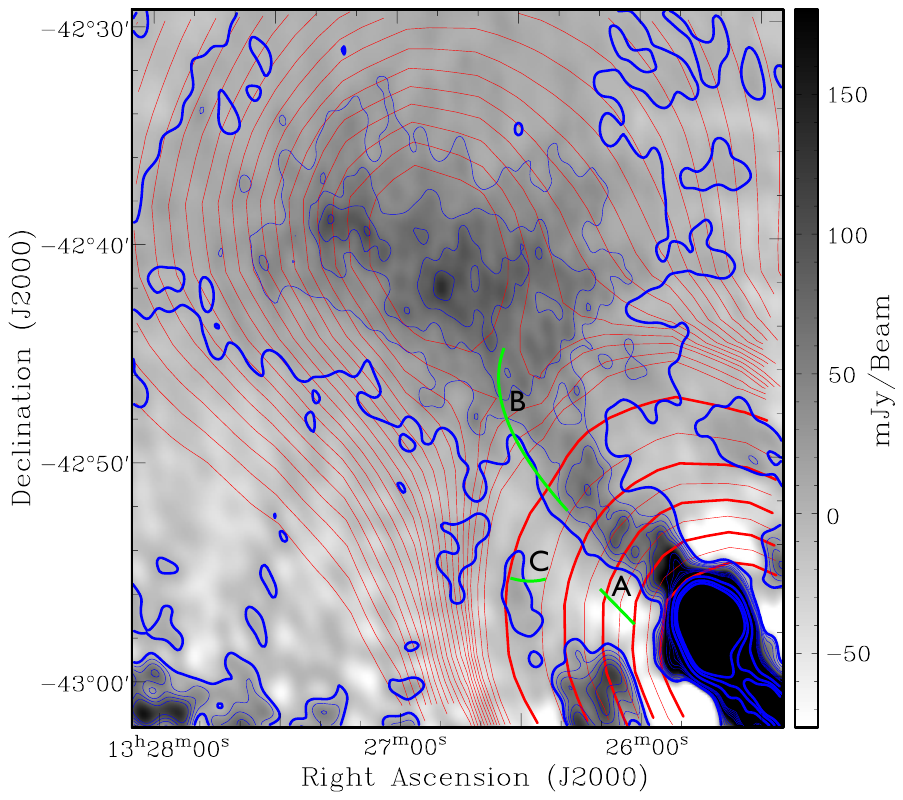} 
         \raggedleft
     \end{minipage}

\caption{Left panel: \citet{morganti1999} image of the NML with the ATCA 1.4~GHz contours (blue) and MWA 154~MHz contours (red) overlaid. Blue contours are at 0, 0.025, 0.05, 0.075, 0.1, 0.15, 0.2, 0.25, 0.3, 0.35, 0.4, 1, 2 and 3 Jy/beam and red contours are at 0.5, 1, 1.5, 2, 2.5, 3, 3.5, 4, 5, 6, 7, 8, 9, 10, 20, 30, 40, 50 and 100 Jy/beam. Right panel: \citet{morganti1999} image of the NML with the ATCA 1.4~GHz contours (blue) and S-PASS 2.3~GHz contours (red) overlaid. Blue contours are at 0, 0.025, 0.05, 0.075, 0.1, 0.15, 0.2, 0.25, 0.3, 0.35, 0.4, 1, 2 and 3 Jy/beam and red contours are at 4 to 13 Jy/beam incrementing in steps of 0.5 Jy/beam and then 14, 15, 20, 30, 40, 50, 60, 70, 80, 90 and 100 Jy/beam. A colour version of this figure is available in the online article.}
\label{morganti}
\end{figure*}

Our radio observations of the northern transition region shed new light on this complex region, which is difficult to image due to its proximity to the extremely bright inner lobes. In Fig.~\ref{morganti} we show the NML region from M99 (image and blue contours), with the MWA 154~MHz contours overlaid in red in the left panel and the Parkes 2.3~GHz contours overlaid in red in the right panel. The MWA and Parkes images have angular resolutions of $\sim$3 and $\sim$11 arcmin, respectively, compared to the higher angular resolutions of the M99 and N15A images of $\sim$50 arcsec. However, these angular resolutions are sufficient to make useful comparisons. The advantages that the MWA has over both the ATCA and the VLA is its excellent instantaneous $(u,v)$ coverage and its abundance of short baselines, which allow us to accurately reconstruct the intermediate scales of the diffuse emission associated with the NML. Our data are well suited to the use of new wide field imaging tools such as \textsc{wsclean} \citep{wsclean}, allowing us to accurately reconstruct diffuse emission without a strong reliance on deconvolution and the unpredictable outputs of imaging algorithms such as \textsc{mem}. 

The MWA 154~MHz and Parkes 2.3~GHz images shown in Fig.~\ref{morganti} confirm the existence of a connection between the NIL and the NML, as first reported by M99. It is possible that the `gap' observed by N15A is an imaging artefact due to one of the many difficulties associated with imaging Cen~A with the VLA, including poor $(u,v)$-sampling at the scales of interest, a very low elevation angle (which amplifies problems associated with the primary beam shape) and contamination from larger-scale emission from the outer lobes. Additionally, the use of \textsc{aips} does not allow for accurate widefield imaging as it cannot account for the curvature of the sky, which starts to become important at the angular scales of interest. The `gap' in the N15A image also runs parallel to other artefacts in their image (see figure 3 of N15A), which could be an indication that it too is not a true physical feature. Another particularly conspicuous feature of the N15A image is the large dark gap on the north-west edge of the NML, which they identify as an imaging artefact. Our image confirms the N15A conclusion that this large dark feature is not real and that the diffuse emission of the NML extends smoothly to the north, gradually fading as it merges into the outer lobe.

Our 154~MHz image does not show any evidence for the extra diffuse emission that appears in the N15A image directly north of the NIL, and connects to the bright ridge of the NML via a large structure that runs parallel to, but north of, the M99 linear connection feature. This structure does not appear in the M99 image either, indicating that this feature is likely to be an artefact of the VLA image. A possible explanation for the differences between the N15A image at 90~cm and the M99 image at 20~cm is the difference in wavelength. Spectral index variations could result in features appearing at one wavelength and being undetectable at another, for example if the bright northerly connection feature of N15A had a much steeper spectral index than the rest of the northern transition region, it may only be detectable in this instance by the VLA at 90~cm. 

Our spectral index results presented in Section~\ref{sec:spectral_index} can be used to determine whether variations in spectral index are a likely explanation for the differences in the images. The spectral index map between 154~MHz and 2.3~GHz (Fig.~\ref{spec_index_figure}) shows that the spectral index of the area corresponding to the connecting jet in the M99 image is approximately $-0.65$, in approximate agreement with the estimate made by N15A. The brightness of the feature in the M99 20~cm image is $\sim$37 mJy/beam, so with a spectral index of $-0.65$, the flux density at 90~cm would be $\sim$98~mJy, making it undetectable in the N15A image, which has an rms noise level of around 50~mJy/beam in that region. This explains the missing feature in the N15A VLA image. The question is then, can the spectral index also explain why the additional diffuse emission seen by N15A does not appear in the M99 image?

Our spectral index map shows that it is not possible that the bright feature in the N15A image to the north-west of the M99 linear jet is missing in the M99 image due to it having a significantly steeper spectral index. The difference in spectral index between the two regions is clearly very small (in Fig.~\ref{spec_index_figure} the north-west region actually appears to have a slightly flatter spectral index). The most likely explanation is therefore that the N15A feature is an imaging artefact i.e. an extra `hook-like' artefact positioned below the similarly-shaped artefacts identified in their figure 3, and the `gap' where the M99 jet should be is simply a non-detection of a faint feature due to the noise level of the image.

\subsubsection{Spectral index between 154~MHz and 1.4~GHz}
\label{sec:mwa_atca_spec}
Having established that the linear feature connecting the NML and the NIL exists, the question remains as to whether this structure is a result of fresh electrons being transported from the AGN in the current epoch, or whether it is a relic of past activity. To investigate this question we have constructed a spectral index map between the MWA image at 154~MHz and the M99 mage at 1.4~GHz (see Fig. \ref{MWA_ATCA_SPEC}). Due to the differences in $(u,v)$ coverage and noise levels between the two images, it is very difficult to construct a meaningful spectral index map under normal circumstances, however, in this instance the  spectral index map can provide an answer to the question of whether the linear feature has a very steep spectral index or not. 

The ATCA image has a minimum baseline length of 31~m and is at a higher frequency, so the minimum angular scale sampled in that image is approximately 23~arcmin. On the other hand, the MWA image, with its abundance of shorter baselines and a lower frequency, samples all angular scales present in the sub-image of the NML. We attempted to match the angular scales of the image by FFT filtering, however were unable to produce a reliable high-pass-filtered MWA image without introducing significant negative imaging artefacts. The best we could do was to remove the mean value of the MWA image. Hence, the spectral indices shown in Fig. \ref{MWA_ATCA_SPEC} represent a lower limit to the true spectral indices as there is clearly missing flux in the ATCA image. This is the reason for the overall steeper spectral index of the NML in Fig. \ref{MWA_ATCA_SPEC} compared to Fig.~\ref{spec_index_figure} and is also the reason for the steepening of the spectral index at the edges of the emission features. 

If the linear connecting feature of the NML is an old relic, no longer supplying energetic particles to the NML, then we would expect a very steep spectral index (e.g. $<-2$) along the linear feature. However, Fig. \ref{MWA_ATCA_SPEC} shows that the feature has a minimum spectral index of $-1$ to $-1.6$ along its length, an indication that fresh electrons are still being injected into the NML from the AGN, in agreement with the conclusions of M99.

\subsubsection{New evidence for a galactic wind}
\label{sec:galactic_wind}

Our MWA image reveals a new radio feature in the northern transition region, which appears not to be a direct result of AGN activity, due to its distance from the large-scale jet and its apparent east-west alignment. The feature is shown in the red MWA contours of the left panel of Fig.~\ref{morganti}, labelled as D, south-east of the inner filament (labelled A). We suggest that this could be evidence supporting the existence of a galactic wind from NGC~5128. Such a galactic wind has been examined in detail by N15B, who found that it could be the driving force behind the `weather system' consisting of bright optical emission line filaments in the northern transition region. N15B go on to show that a galactic wind with energy supplied by both starburst activity and the AGN and interacting with a ring of atomic hydrogen (\citealt{HI1994}, \citealt{struve2010}), can explain many of the features present in this complex region of Cen~A. 

N15B argue that their non-detection of a large-scale radio jet connecting the NIL and NML is evidence that a galactic wind must be playing a significant role in powering the weather system, which would otherwise fade or dissipate within a few Myr (N15B), but they do not find any direct evidence for the galactic wind in their radio images. The radio feature seen in our MWA image could be direct evidence of such a wind, forming an expanding bubble that is coming into contact with the atomic hydrogen to the north of the galaxy. This claim is supported further by the presence of an additional bright emission-line filament that has not previously been discussed in the literature, which coincides with the northern boundary of the new radio feature and is clearly seen in our emission line images to be discussed in Section~\ref{sec:emission_lines}. It is also expected that a galactic wind bubble would coincide with X-ray emission (N15B) and it can be seen in figure 5 (left panel) of N15B that there is indeed some X-ray emission \citep{kraft2009} present at the location of our newly identified emission-line filament at RA~(J2000)~13\textsuperscript{h}26\textsuperscript{m}30\textsuperscript{s}, Dec~(J2000)~$-42$\degr 56$'$.

N15B postulate that the galactic wind in Cen~A, which would normally propagate as a pair of roughly spherical bubbles on each side of the galaxy, is likely to have been disturbed by the radio jet and the cocoon, which has already been excavated by the radio plasma \citep{kraft2009}, resulting in the galactic wind being able to propagate out faster in the location and direction of the AGN jets. Our radio observations support this scenario, where we can see an additional radio bubble expanding into the ISM in a much broader pattern than the jet connecting the NIL and NML (see left panel of Fig~\ref{morganti}, red contours surrounding label D). At its furthest extent this bubble is approximately 10~kpc from the centre of NGC~5128, in approximate agreement with the N15B calculations for a spherically expanding bubble, which would have a radius of 7~kpc without the presence of the radio cocoon, and 13~kpc for an `AGN-enhanced' wind. It is possible that the galactic wind bubble from NGC~5128 that we see in the radio emission could be of similar origin to the large bipolar outflows from the centre of our own Galaxy \citep{carretti2013}, which extend 8~kpc above and below the Galactic plane and coincide with the \emph{Fermi} bubbles observed in $\gamma$-rays \citep{su2010}. 


\subsection{Optical emission lines in the NML: photoionisation from star formation and the AGN}
\label{sec:emission_lines}

In our H$\alpha$ image of the NML region (Fig. \ref{optical_emission_lines}) we can clearly see the `inner filament', the `outer filament' and a previously unreported filament further to the east. The locations of these filaments are also shown as green lines in Figs~\ref{morganti} and \ref{MWA_OPTICAL_OVERLAY} and are labelled A, B and C, respectively. 

Focusing on the outer filament (since this is where the MMTF is tuned to), we measure a total H$\alpha$ flux of $f($H$\alpha)=6.4 \times 10^{-13}$ erg s$^{-1}$ cm$^{-2}$ over the whole filament, which corresponds to a star-formation rate (SFR) of $7 \times 10^{-3} \ M_{\sun} \ $yr$^{-1}$. The total H$\alpha$ flux was summed from a quadrilateral aperture containing the filament, which was increased in size until the total flux reached a constant value. The SFR was calculated from the continuum-subtracted H$\alpha$ emission using the \citet{kennicutt1998a} relation. In comparison, for part of the outer filament \citet{mou00} found $f($H$\alpha)=1.3 \times 10^{-13}$ erg s$^{-1}$ cm$^{-2}$ and SFR$=1.2 \times 10^{-3}\ M_{\sun} \ $yr$^{-1}$. For the same part of the filament as \citet{mou00}, we find similar numbers of $f($H$\alpha)=1.2 \times 10^{-13}$ erg s$^{-1}$ cm$^{-2}$ and SFR$=1.3 \times 10^{-3}\ M_{\sun} \ $yr$^{-1}$. The area over which the outer filament flux was measured was 34,000 arcsec$^{2}$, so the average H$\alpha$ surface brightness is $1.87 \times 10^{-17}$ s$^{-1}$ cm$^{-2}$ arcsec$^{-2}$. 

Another important quantity is the emission-line ratio [N\textsc{II}]/H$\alpha$, which we measure only for the diffuse, low surface brightness region in the centre of the outer filament (due to instrumental effects of the MMTF mentioned in Section \ref{sec:MMTF_section}). At this location in the filament we measure [N\textsc{II}]/H$\alpha = 0.23 \pm 0.01$, indicating a relatively low level of ionisation.

\citet{salome2016} have calculated the SFR in the outer filament of Cen~A using three tracers; FUV and TIR (using SFR relations described by \citealt{kennicutt2012}) and FIR (using SFR relations described by \citealt{kennicutt1998b}). Our SFR calculated from the H$\alpha$ flux of $7 \times 10^{-3} \ M_{\sun} \ $yr$^{-1}$ is only slightly higher than SFRs calculated by \citet{salome2016} of 1.4, 4.9 and 5.5 $\times 10^{-3}\ M_{\sun} \ $yr$^{-1}$ for FUV, TIR and FIR, respectively. It has been found that the inner and outer optical filaments are sites containing young stars \citep{rej2002} and that this could be an example of jet-induced star formation, especially in the inner filament where the position angles of the AGN jet and the filamentary structure are well-aligned \citep{vanB1993,dey1997,graham1998,mou00,crockett2012,hamer2015}. However, it has also been found that the star formation in these regions is insufficient to explain the high levels of ionisation in the clouds that make up the bright filaments and that additional ultraviolet photons originating from the AGN must be contributing to the formation and continued supply of energy to maintain the filaments \citep{santoro2015a,santoro2015b,santoro2016,rej2002,morganti1991,morganti1992}. 

Indeed, strong evidence for an `ionisation cone' originating from the Cen~A AGN has been found by \citet{sharp2010}, consistent with the picture that the AGN is supplying ionising photons to the filaments, well out into the weather system of the northern transition region of Cen~A (\citealt{santoro2015b}, \citealt{hamer2015}, N15B). To assess the likelihood that the influence of the ionisation cone extends out into the transition regions of Cen~A it is instructive to examine our nearest neighboring supermassive black hole Sgr~A$^{\star}$.

\citet{JBH2013} have proposed that an excess of H$\alpha$ emission in the Milky Way's Magellanic Stream can be explained by a relatively recent `Seyfert flare' event in our own Galaxy. This is an interesting comparison to make, since the calculations of \citet{JBH2013} can be directly applied to the case of the outer filament of Cen~A. For the Magellanic Stream, \citet{JBH2013} assume an Eddington fraction of $f_E=0.1$ for Sgr~A$^{\star}$ and a black hole mass of $4 \times 10^{6} \ M_{\sun}$. For Cen~A, the black hole mass is an order of magnitude higher at $5.5 \times 10^{7} \ M_{\sun}$ \citep{cappellari2009}, but the Eddington fraction is lower, estimated at $f_E=0.01$ \citep{beckmann2011}. These two factors cancel each other out in the calculation of the expected H$\alpha$ surface brightness (see equations 11-13 of \citealt{JBH2013}). Rescaling to a distance of 15~kpc for the outer filament of Cen~A (\citealt{JBH2013} use a distance of 55~kpc to the Magellanic Stream), we find an expected H$\alpha$ surface brightness at the peak of the flare of $7 \times 10^{-17}$ s$^{-1}$ cm$^{-2}$ arcsec$^{-2}$. This is in excess of the observed H$\alpha$ surface brightness by a factor of almost 4, which allows for the fading of the H$\alpha$ in the intervening time between the AGN flare and the observation.

It is possible that other phenomena are also contributing to the photoionisation of the filaments. For example, it has been suggested by \citet{sutherland1993} that shocks resulting from jet-ISM interactions could produce the emission-line filaments in Cen~A. However, \citet{santoro2015a} show that while the kinematics of the gas do indicate a jet-ISM interaction, it is a relatively soft interaction, lacking the fast shocks necessary to fully explain the high ionisation of the gas. Another possible energy source for the ionisation of the filaments is thermal conduction as proposed by \citet{sparks2004} for the filaments in M87. In the case of Cen~A, however, this appears unlikely as the hot X-ray emission in the NML \citep{kraft2009} is not spatially coincident with the optical filaments.

The `new' filament (which is also visible in the FUV image of N15B fig 1, but not discussed in the text), is oriented almost perpendicular to the position angle of the AGN jets and runs roughly parallel with the northern edge of the new radio feature discussed in Section~\ref{sec:galactic_wind}. We suggest that this feature could be the result of a galactic wind. In this case, the galactic wind could be supplying ionising photons to energise this filament, which marks a point where the resulting radio bubble is coming into contact with cold HI to produce weather features similar to those seen in the inner and outer filaments and analysed in detail by N15B. 

The new filament is mostly a diffuse, low surface brightness feature similar to the central region of the outer filament where we measured the [N\textsc{II}]/H$\alpha$ ratio and found a relatively low level of ionisation. If this new filament has a similarly low level of ionisation it could be consistent with ionisation from a galactic wind. Further observations of this region are required to determine the level of ionisation in the gas, the ages of the stars within the filament, and the relative contribution that in-situ star formation is making to the ionisation of the gas. Comparisons between NGC~5128 and ionisation due to galactic winds in other galaxies \citep{v2003} and our own Galaxy (\citealt{JBH2003}, \citealt{JBH2013}) may also help to reveal the physical mechanisms responsible for the formation and evolution of this ionised filamentary structure.

Throughout the above discussion we have assumed a bipolar galactic wind and AGN jets, but not mentioned the lack of any corresponding bright radio features (apart from the radio polarisation features seen by \citealt{osullivan2013}) or ionised weather systems to the south of NGC~5128. Our radio observations agree with all previous observations of Cen~A, including N15A, in that we do not detect a southern counterpart to the NML in our total intensity maps, which supports the hypothesis of N15B and others that the lack of radio emission and complex weather in the southern transition region is related to the distribution of HI in a ring around NGC~5128. In this case, the southern AGN jet and galactic wind bubble are not coming into contact with any cold HI gas and the HI cloud seen to the south of the galaxy \citep{HI1994,struve2010} is sitting out of the plane of the AGN jet in three dimensions. The gas-ISM-AGN-galactic wind interactions are complex phenomena common to galaxies throughout the Universe (N15B, \citealt{morganti2016}, \citealt{santoro2015a}, \citealt{hamer2015}, \citealt{morganti2010}, \citealt{v2005}) and further study of our nearest neighbouring active galaxy, Cen~A, will enable a better understanding of the astrophysics occurring as galaxies interact with their environments.

\subsection{Halo stars of NGC~5128: An old population affected by a recent merger and AGN activity}
\label{sec:halo}
\begin{figure*}
\centering 
\includegraphics[clip,trim=50 150 50 150,width=0.9\textwidth,angle=270]{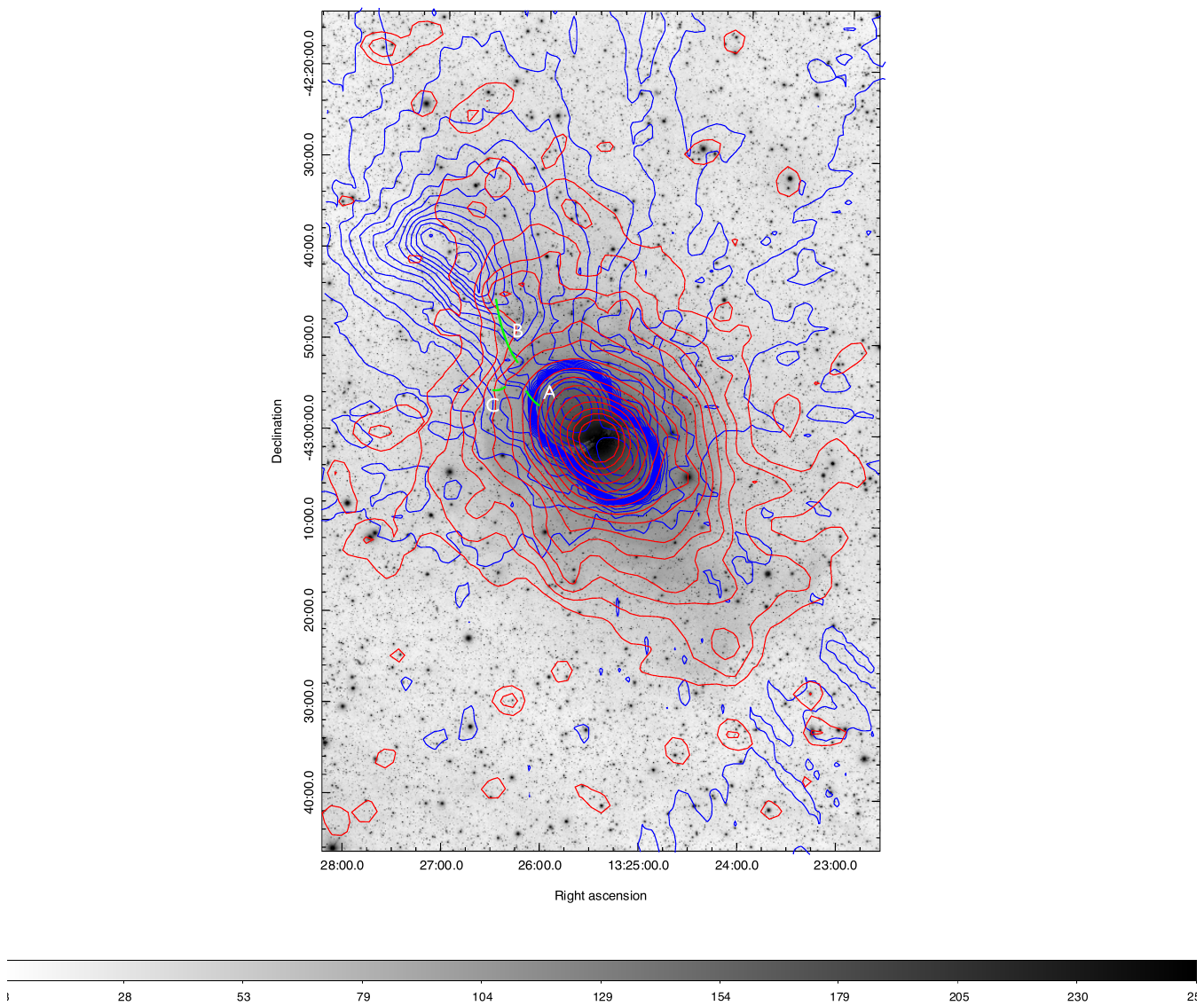}
\caption{Deep optical image overlaid with optical image smoothed to MWA resolution (red contours from 20$\%$ to 90$\%$ of maximum value, incrementing in steps of 5$\%$) and overlaid with MWA 154~MHz contours (blue contours from 0.5 to 10 Jy/beam incrementing in steps of 0.5 Jy/beam and then 16 to 256 Jy/beam incrementing by a factor of two). The positions and extents of the optical emission line filaments are shown in green and labelled A for the inner filament, B for the outer filament and C for the newly detected filament to the east. A colour version of this figure is available in the online article.}
\label{MWA_OPTICAL_OVERLAY}
\end{figure*}

Our deep optical image shown in Fig.~\ref{optical1} clearly shows the concentric shell structure first revealed by \citet{malin1983} and subsequently imaged by \citet{peng2002}, without the need for any image post-processing, such as unsharp masking, to accentuate the shells. This shell structure is evidence of the merger that occurred some 10$^8$ years ago (\citealt{malin1983}, \citealt{quinn1984}, \citealt{charmandaris2000}) and the number and positions of the shells in our image are in agreement with \citet{malin1983} and their findings. Of more interest for this particular study is the diffuse extension of optical emission to the north-east and south-west of NGC~5128 (first identified by \citealt{johnson1963} and subsequently imaged by \citealt{cannon1981} and resolved into individual stars by \citealt{crn2016}), which is misaligned with both the current AGN jets and inner radio lobes and the NML. This misalignment is shown clearly in Fig.~\ref{MWA_OPTICAL_OVERLAY}, where we have overlaid our deep optical image with the MWA 154~MHz contours in blue and the contours of the deep optical image, smoothed to the MWA angular resolution, in red. 

It is known that the stars in the outer halo of NGC~5128 (in both the north and the south) are, for the most part, on the order 11-12~Gyr old (\citealt{rej2005}, \citealt{rej2011,crn2016}). This is far older than even the largest estimates for the ages of the outer radio lobes ($\sim$1Gyr), indicating that the majority of the stars were not formed as a result of jet-induced star formation (with the exception of the stars within the previously discussed emission-line filaments). Jet-induced star formation then does not explain the bipolar structures we see in our deep optical image. In their examination of these faint diffuse extensions to the halo of NGC~5128, \citet{cannon1981} suggest that the emission could be from stars or reflected starlight from dust. The density map of red giant branch stars by \citep{crn2016} clearly shows that the stellar density profile is extended along the direction of the optical extensions our deep optical image. 

While NGC~5128 is classified as an E/S0 galaxy, at the low-end of stellar density distribution out in the extended halo, its ellipticity is much more pronounced and has a minor axis misaligned by almost $90\degr$ to the spin axis of the galaxy, which is unusual since even for elliptical galaxies alignment with the spin axis is preferred \citep{chisari2017}. It is difficult to determine from the available data and simulations whether this ellipticity and misalignment is due to the merger process or if the AGN has had some influence. If some of the optical emission is reflected starlight from material that has been entrained by the expanding radio plasma, this material would have initially tracked a more north-south path, but has drifted towards the orientation of the NML following the precession \citep{haynes1983} or re-orientation of the AGN jets. Thus resulting in starlight-reflecting dust positioned in a bipolar structure with a position angle in-between that of the outer lobe and the inner lobe in the north, and diametrically opposed in the south. In this scenario, the optical extensions and unusual shape and orientation of the extended halo of NGC~5128 could be evidence of kinetic feedback related to the AGN and operating over long timescales.

\subsection{The missing Southern Middle Lobe: AGN interactions with a HI ring}
\label{sec:HI}

In this discussion we have attributed the asymmetry of Cen~A on intermediate scales (i.e. the lack of a southern middle radio lobe and a southern transition region  weather system) to the lack of cold HI in direct contact with the AGN outflow and galactic wind to the south. This lack of interaction between outflows from the central galaxy and cold gas may also explain the asymmetry that is seen on large scales. In the south, the brightest part of the outer lobe begins at about twice the distance from corresponding distant from the core that the NML ends in the north. At the corresponding point in the northern outer lobe there is a distinct lack of emission and instead, the bright NML changes its orientation and gradually fades to the north-west. This `missing' part of the northern outer lobe could be the result of the flow of plasma from the AGN, out through the large-scale jet, being interrupted by the cold HI ring and creating the NML in place of the corresponding bright feature seen in the southern outer lobe. Interestingly, \citet{osullivan2013}, find that the fractional polarisation of the radio emission at the expected location of the southern middle lobe is similar to that of the NML and suggest that the southern middle lobe may have dissipated after mixing with a large amount of thermal gas from the host galaxy. Future polarisation studies with the MWA may shed further light on this scenario, particularly with the large $\lambda^2$-coverage provided by the MWA.

\subsection{Cen~A as a local laboratory for AGN feedback mechanisms}
\label{sec:feedback}

In this work we see two modes of AGN feedback in action through the photoionisation of gas and the outflows of material via jets and winds. There appears to be a complex interrelationship between AGN jets and galactic winds, with feedback occurring through both narrow and wider solid angles. In current cosmological simulations, AGN feedback is implemented on a scale of order the resolution of the simulations, which is probably larger than the actual physical scale, and even the most advanced simulations cannot match the observed properties of galaxies very well \citep{nelson2015,weinberger2017}. As the resolution of cosmological simulations improves, a detailed understanding of AGN feedback mechanisms will be required and observations of Cen~A will play a crucial role in determining how these feedback process can be most accurately incorporated to reproduce the properties of galaxies that we observe throughout the Universe.

\section{Conclusions}

We have taken advantage of the widefield observational capabilities of radio and optical instruments to investigate the properties of the Cen~A `transition regions' at intermediate angular scales, providing new insights into the interactions occurring between the AGN and the surrounding environment. We have presented new low-frequency observations of Cen~A at 154~MHz from the MWA and 2.3~GHz from the Parkes S-PASS survey. We have also presented new optical emission line images from the MMTF on the Magellan Telescope and the deepest available optical images from amateur observations. Our new widefield observations have enabled us to make the following conclusions: \\
1. The spectral index of the outer lobes is remarkably uniform, with little variation along the major axis or in the transverse direction, indicating that energy is being continually supplied from the AGN and in-situ particle acceleration is taking place throughout the lobes.\\
2. There \emph{is} a collimated structure, a `large-scale jet' connecting the NML and the NIL, as first observed by \citet{morganti1999}.\\
3. Our low frequency radio image shows evidence for a galactic wind bubble protruding to the west of the NIL, possibly interacting with the HI ring to produce X-ray emission.\\
4. We have identified a new emission-line filament (in addition to the well-studied `inner' and `outer' filaments), which may be further evidence of an interaction between a galactic wind from NGC5128 and HI in the surrounding ISM. \\
5. The outer optical halo of NGC~5128 is elongated in the north-east and south-west directions, but this elongation is misaligned with the NML and the bright optical filaments. This hints at kinetic feedback provided by a bipolar outflow from the AGN. \\
6. We see no evidence for a southern counterpart to the NML in our radio images and attribute this to there being a lack of cold gas in contact with the southern outflow from the AGN in the expected position of a `southern middle lobe'.\\

We have also explored similarities between structures in our own Galaxy and Cen~A. In the Milky Way we can resolve complex structures, hidden in other galaxies due to resolution limitations. While in Cen~A, we are able to image structures that are physically much larger and are difficult to image in our own Galaxy (for example, even for wide-field-of-view radio interferometers such as the MWA, the 50\degr \ radio lobes of the Milky Way are too big to image), while still being able to compare the radio morphology to detailed images of optical filaments and $\gamma$-ray data. More detailed radio observations of the northern transition region of Cen~A, including a detailed spectral-index analysis, will be required to confirm the existence of the galactic wind structure and determine its true nature. We plan to conduct new observations of Cen~A with the MWA in its new, long-baseline configuration, which will double the angular resolution of the telescope. These new observations may provide data crucial as inputs to high-resolution cosmological simulations in the future, which will be able to resolve the energy transportation due to the different modes of AGN feedback.


\section*{Acknowledgements}

This scientific work makes use of the Murchison Radio-astronomy Observatory, operated by CSIRO. We acknowledge the Wajarri Yamatji people as the traditional owners of the Observatory site. Support for the operation of the MWA is provided by the Australian Government (NCRIS), under a contract to Curtin University administered by Astronomy Australia Limited. Parts of this research were conducted by the Australian Research Council Centre of Excellence for All-sky Astrophysics (CAASTRO), through project number CE110001020. The Parkes Radio Telescope is part of the Australia Telescope National Facility, which is funded by the Commonwealth of Australia for operation as a National Facility managed by CSIRO. This work has been carried out in the framework of the S-band Polarization All Sky Survey (S-PASS) collaboration. We acknowledge the Pawsey Supercomputing Centre, which is supported by the Western Australian and Australian Governments. We also acknowledge funding from NSF grant AST/ATI 0242860 and NSF contracts AST 0606932 and 1009583. JBH is funded by a Laureate Fellowship from the ARC. RM gratefully acknowledges support from the European Research Council under the European Union's Seventh Framework Programme (FP/2007-2013)/ERC Advanced Grant RADIOLIFE-320745. BM is funded by a Discovery Early Career Researcher grant from the ARC. We acknowledge that the University of Melbourne Parkville campus, upon which most of this paper was written, is located on the land of the Wurundjeri people.







\bsp	
\label{lastpage}
\end{document}